# Resource Management Techniques for Cloud-Based IoT Environment

**Syed Arshad Ali, Manzoor Ansari and Mansaf Alam**

Department of Computer Science, Jamia Millia Islamia, New Delhi, India

arshad158931@st.jmi.ac.in, manzoor188469@st.jmi.ac.in, malam2@jmi.ac.in

**Abstract** Internet of Things (IoT) is an Internet-based environment of connected devices and applications. IoT creates an environment where physical de- vices and sensors are flawlessly combined into information nodes to deliver innovative and smart services for human-being to make their life easier and more efficient. The main objective of the IoT devices-network is to generate data, which are converted into useful information by the data analysis process, it also provides useful resources to the end-users. IoT resource management is a key challenge to ensure the quality of end-user's experience. Many IoT smart devices and technologies like sensors, actuators, RFID, UMTS, 3G, and GSM etc. are used to develop IoT networks. Cloud Computing plays an important role in these net- works deployment by providing physical resources as virtualized resources consist of memory, computation power, network bandwidth, virtualized system and device drivers in secure and pay as per use basis. One of the major concerns of Cloud-based IoT environment is resource management, which ensures efficient resource utilization, load balancing, reduce SLA violation, and improve the system performance by reducing operational cost and energy consumption. Many researchers have been proposed IoT based resource management techniques. The focus of this paper is to investigate these proposed resource allocation techniques and finds which parameters must be considered for improvement in resource allocation for IoT networks. Further, this paper also uncovered challenges and issues of Cloud-based resource allocation for IoT environment.

# 1 Introduction

The Internet of Things (IoT) is a set of connected smart device and sensors over the Internet. These devices are connected using the wired/wireless network technologies to communicate and transfer data from one node to another [58, 28, 4, 46]. The things in IoT infrastructure network are sensors, smart devices, sensor data, software agents and human beings [57]. These networked independent devices make local network and connected to the global network to share information with others in real time to realize that the Things are connected into the real world and connect all devices [5, 34, 41]. In cyber-physical ecosystem, each edge-node is supposed to an IoT device which can dynamically cooperate with other devices in the network to execute one or more user's tasks allocated to the system network. Resources like processing power, storage, network bandwidth, RAM are usually limited in these IoT network, though these infrastructures and computing resources are provided by the Cloud service providers. IoT devices produce huge amount of real-time data from the sensors. Cloud storage is used for storing these real-time data on different local networked data centers, which upload these data to the global networked data centers to allow access for all globally situated smart devices [15]. In this paper, the author studied various resource allocation techniques for Cloud-based IoT system. Classification of these techniques have been done based on parameters like QoS, context, cost, energy consumption and SLA. Furthermore, the author also discussed various parameters of resource allocation techniques of IoT system.

# 2 Basic Concepts of IoT, Cloud and Resource Management

## 2.1 *What are the basic elements of IoT environment?*

Internet of things offers numerous advantages and services to the users. Therefore, to use them correctly, some elements are needed. The IoT elements will be discussed in this section. Figure 1 shows the elements required to provide IoT functionalities.

**Identifiers**

Within the network, it offers an explicit identity for each object. In identification, two processes exist naming and addressing. The naming relates to the object's title whereas the addressing explains address of an object. These two processes are very different even though two or more objects could have the same name, but they are always different and unique. There are several approaches accessible which expedite the naming of objects in the network, such as ubiquitous codes (uCode) and electronic product codes (EPCs) [32]. Using IPv6, each object has a unique address. First, IPv4 was used to allocate the address, but due to a huge amount of IoT devices, it was unable to address the need. IPv6 is therefore used because it uses a 128-bit addressing system.

**Sensing Devices**

This involves acquiring information from the environment and transferring it to a local, remote or Cloud-based database as an instance of the IoT sensor. We can identify intelligent devices, portable sensors or actuators. The collected information is transmitted to the storage medium. Numerous detection devices to collect data on objects such as RFID tags, actuators, portable sensors, smart sensors, etc.

**Communications Devices**

To achieve smart services, IoT communication techniques communicate heterogeneous artifacts. One of the main goals of Internet of things is Communication in which various devices connect and communicate with each other. In the communication layer, devices can transfer and deliver messages, documents and other information. There are many methods which facilitate communication, for example Bluetooth [42], radio frequency identification (RFID) [1], long term evolution (LTE) [21], Wi-Fi [25] and near-field communication (NFC) [55].

**Compute Devices**

The computation of the data collected by the objects is achieved using sensors. It is used to develop processing in IoT applications. Raspberry Pi, Arduino, and Gadgeteer are utilized for hardware platforms, whereas the operating system plays a significant role in the processing of software platforms. There are various kind of operating systems are used, including Lite OS [16], Riot OS [14], Android, Tiny OS [35] etc.

**Services IoT**

Applications provide four types of services [56] [26]. The first service that is associated with an identity. It is used to acquire the identity of the objects which sent the request. The aggregate of information is another service aimed at collecting all the data about the objects. The aggregation service also performs the processing. The third service refers to co-operative service that makes decisions-based on the information gathered and transfer suitable rejoinders to the devices. The last service is the pervasive service, which is used to replace devices immediately without rigidity in terms of time and place.

**Semantics**

They are the IoT's concern to facilitate the consumers who perform their tasks. To fulfill its responsibilities, it is the most important component of IoT. It performs as the IoT's brain. It accepts all the information and makes the appropriate decisions to send responses to the devices.

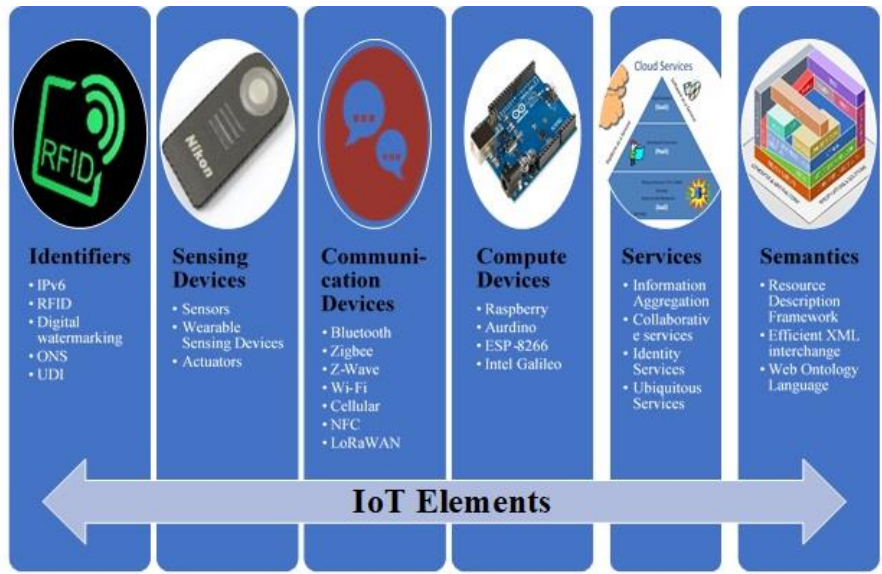

**Fig. 1.** Basic elements of IoT environment

The key techniques used in each IoT component are presented in Table 1.

**Table 1.** Key Technologies used by IoT Components

| IoT components | Main Technologies |
| --- | --- |
| **Identification** | IPv4, IPv6 Electronic product code(eCodes), ubiquitous code (uCode), |
| **Sensing** | Actuators, Sensors, Wearable Sensing Devices, RFID Tags. |
| **Communications** | Wireless Sensor Network (WSN), Long Term Evolution (LTE), Bluetooth, Near Field Communication (NFC), Radio Frequency Identification (RFID), |
| **Computation** | Intel Galil Operating System, Arduino, Raspberry Pi. |
| **Services** | Collaborative-Aware, Ubiquitous Identity-Related, Information Aggregation |
| **Semantics** | EXI, OWL, RDF |

## 2.2  What are the various IoT architecture frameworks?

An IoT architecture can be viewed from three viewpoints: Things-oriented, Internet- oriented and semantic-oriented [12]. In things-oriented viewpoint the intelligent autonomous smart devices are connected to each other using NFC and RFID technologies for specific daily life applications. Internet-oriented viewpoint focuses on how these smart devices connect to the internet using unique identification (IP addresses) and standard communication protocols to facilitate the global connections among these applications based smart devices. In semantic viewpoint of IoT architecture, the data generated by the IoT devices are used to generate useful information and handle the architectural modeling problems efficiently using these produced information [40].

In the opinion of most researchers on conventional IoT architecture, it is considered at three layers: -Perception Layer, Network Layer, Application Layer. In addition, some researchers have analyzed another layer also included in the latest IoT architecture, which is a support layer located between the network layer and the application layer. Multi-tier architecture of IoT is displayed in Figure 2.

The support layer consists of fog computing and Cloud Computing. In this section, we define seven-tier architecture: collaboration and processes, applications, data abstraction, data accumulation, edge computing, connectivity, physical devices and controller. The basic layered architecture of IoT and its components in each layer have been depicted in Figure 3.

**Layer 1: Physical Devices and Controllers**

It is a layer of perception or hardware that collects and sends information from the physical world to the next layer. This layer includes objects and physical sensors. Basically, this layer is intended to detect various objects and collect environmental data such as humidity, temperature, water quality, pressure, air quality, motion detection etc. Controllers and Physical devices can control multiple devices. These are the "things" in the IoT containing a wide range of endpoint devices for sending and receiving information. The list of devices is already extensive today. It will be effectively unlimited as more devices will be added to the IoT over time.

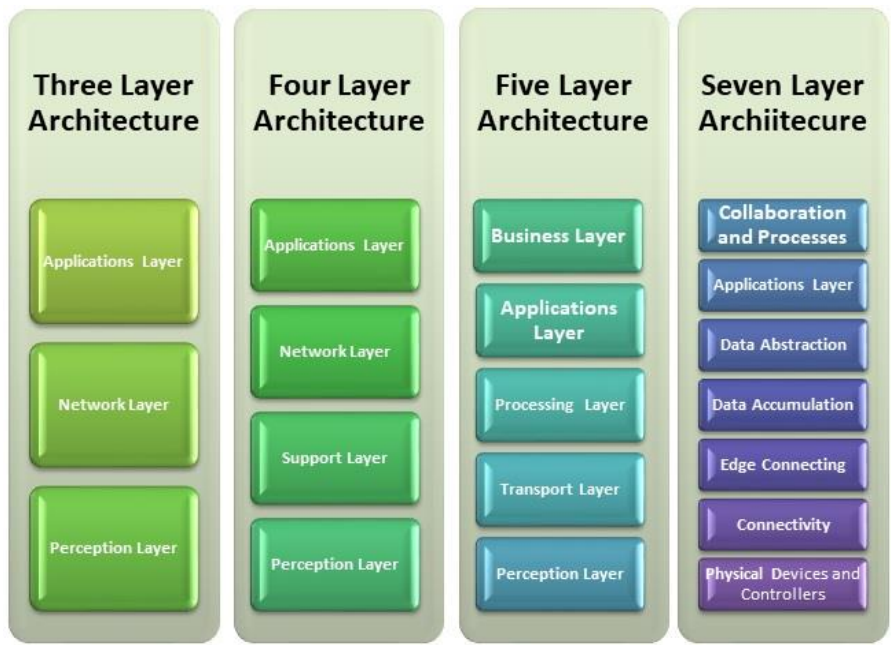

**Fig. 2.** Multi-Tier IoT Architecture

**Layer 2: Connectivity**

This layer is used to interact with various IoT components via interconnecting systems such as switches, gateways as well as routers. In addition, it transfers data collected strongly from the sensors to the top layer for processing. It includes transferring data from physical devices to Cloud or other devices, which may be in the form of ZWave, SigFox Zigbee or Bluetooth. This layer extends to Cloud transport services from the "intermediate" of an Edge Node device.

**Layer 3: Edge Connecting**

The next phase is Edge Computing, or more appropriately "Cloud Edge" or "Cloud Gateway". Layer 3 requires data from the connectivity phase and makes it suitable for archiving and higher-level processing information. The processing elements in this layer work with a large volume of data that could transform some facts to moderate data size.

**Layer 4: Data Accumulation**

They must be stored after the data is accumulated. It is important where information is stored. While some data may be stored to the limit, most data will have to be delivered to the Cloud. Big Data machines will be able to exploit their computing power and analyze the data. The main objective of this layer is to store the data of Phase 3. Acquire and store a large amount of data and place them in the warehouses so that they are accessible from the upper layers. As a result, it simply modifies event-based data in query-based processing information for higher layer. This layer can be deployed in SQL or requires a more sophisticated Hadoop and Hadoop file system, Mongo, Cassandra, Spark or other NoSQL solutions.

**Layer 5: Data Abstraction**

This layer combines data from different sources and converts the stored data to the appropriate application format in a manageable and efficient way [11]. A main component of the large-scale high-performance implementation architecture is a publishing/subscription software framework or a data distribution service (DDS) to simplify data movement between edge computing, data accumulation, application layers and processes. Whether it's a high-performance service or a simple message bus, this infrastructure simplifies deployment and improves performance for all applications, except for the simplest.

**Layer 6: The Application Layer**

It applies to information elucidation from different IoT applications. It comprises several IoT applications, like medical care, smart city, smart network, smart agriculture, smart building, connected car etc. [51]. This phase is self-explanatory where the application logic of the control plan and the data plan are performed.

Process optimization, logistics, statistical analysis, control logic, Monitoring, alarm management, consumption models are just certain cases of IoT applications.

**Layer 7: Collaboration and Processes**

At this phase, application processing and collaboration are presented to users, and the processed data in the lower layers is incorporated into commercial purposes. It encompasses collaboration, people, businesses, and decision-making processes based on IoT-derived information. This layer classifies people who can collaborate and communicate to use IoT data proficiently. It delivers additional features, such as the creation of commercial graphics and models and other data-based recoveries from the application layer. It also helps executives make accurate business decisions based on data analysis [44].

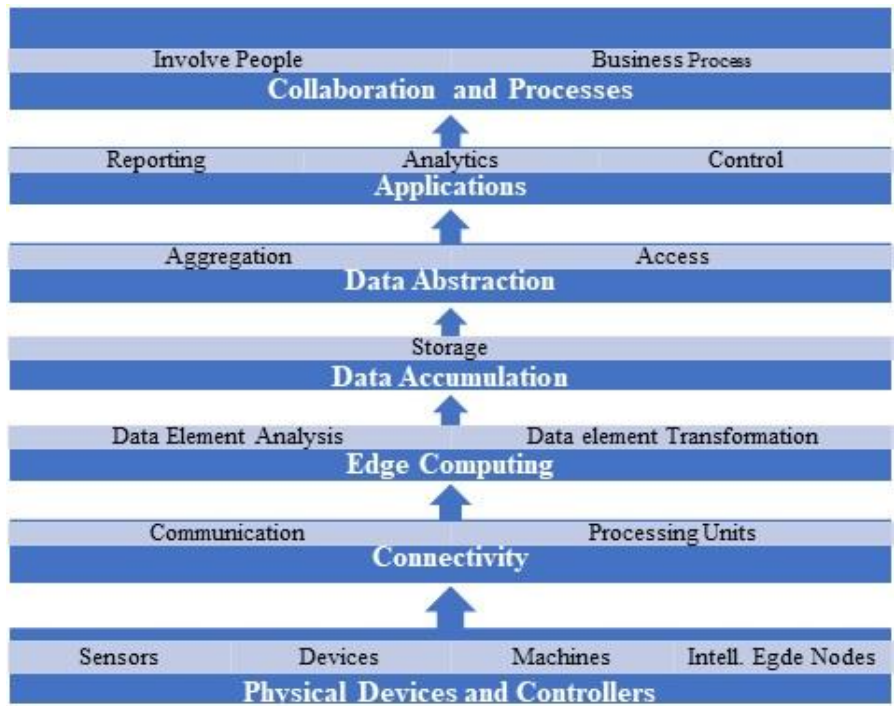

**Fig. 3.** Basic Architecture of IoT

## *2.3 How Cloud Computing supports IoT Infrastructure?*

National Institute of Standards and Technology (NIST) defined, Cloud Computing as a paradigm that enables global, appropriate and on-demand self-serviced shared pool of computing resources (network, storage, applications and services) to the end-user on pay per use basis with minimal effort or service provider communication [43]. Cloud Computing is used rapidly by IT industries and professional for deployment of their projects due to minimal cost and rapid elastic characteristics of Cloud- services [53]. Cloud Computing is an Internet based technology that facilitates both service-consumers and providers by its essential features of On-demand self-service, broad network access, resource pooling, rapid elasticity and measured services [59]. Cloud Computing has three service models and four deployment models. Cloud Service provider's software running on Cloud set-up is used by the Cloud consumers in Soft- ware as a service (SaaS) model. Cloud consumers can deploy their applications on to the provider's infrastructure using software design languages, libraries and tool provided by the Cloud infrastructure in Platform as a Service (PaaS) model. In Infrastructure as a Service (IaaS) model, consumer can provision pol of resources like storage, processing elements, network and other basic computing requirement for running any software and operating system provided by the Cloud provider as a virtual machine (VM) [54]. Cloud Computing also provide four deployment model as per the user priorities and features named as public, private, community and hybrid Cloud deployment models [52]. Due to massive features and ability, Cloud Computing is used by many growing technologies, IoT is one among them. Integration of IoT with Cloud Computing has been intensively used by many real-life applications like smart cities, healthcare, agriculture industries, transportation, smart vehicles and many more [38]. The current trends of technologies are moving towards the use of globally connected smart devices, which produces huge amount of data that cannot be stored in locally due to limited storage capacity. The running fuel of all industries is the data, data is useless without its analytics to get the useful information for industries future-plans and policies. The huge amount of data gathered by the smart connected devices needs powered computing systems that should capable enough to compute this huge amount of data. The local systems are not capable enough to process these data for analytics. These limitations of huge storage space, computing power and network bandwidth can be overcome using pooled resources provide by Cloud Computing. Figure 4 illustrates how Cloud Computing supports IoT for smooth functionality. Cloud Computing infrastructure is used between the applications and IoT devices as a hidden layer, where all functionalities are hidden from the implementation of IoT based applications [24].

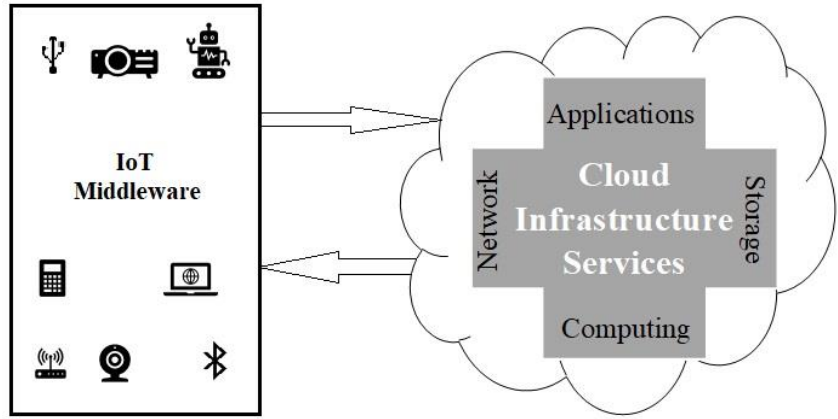

**Fig. 4.** How Cloud Support IoT Applications

## 2.4 Why Resource Allocation is important for IoT?

Highly efficient, maintained and cost-benefit network ensure the Quality of Service (QoS) standards. IoT architecture has various resources connected with the network. The resource allocation is an important aspect for QoS standards because of efficient and effective allocation of resources in the IoT network. Resource allocation is also responsible for high standard of security because in IoT architecture the data is divided in many data streams gathered from different sensors and different types of services are provided by the networked devices. The IoT networked resources consists computing elements, storage and energy. Efficient Cloud resource allocation helps IoT networked devices to utilize these resources in an efficient and cost-effective way to improve system performance and productivity. IoT devices and resources are heterogeneous and globally distributed in nature, therefore, resource allocation and management are very important aspect of IoT environment. The entity in the IoT network system can be an object, a human being or a place that is used to communicate between the IoT system application and the system user. These allocated objects are known as resources which can be categorized based on the information they communicate in the network. The three layered IoT system mainly used are classified in IoT things, edge and Cloud infrastructure as described in Figure-5. Allocation of resources by the system to accomplish the user tasks required various phases, first the system selects the required resource from these three layers, then it selects the nodes which are capable to execute the user's tasks and after that it schedule the user's task to the selected node for its execution. Finally, the communication among these networked resources are done to accomplish the successful execution of the user's task. Resource allocation has many aspects of resource discovery, resource provisioning, resource scheduling and resource monitoring as well. An effective Resource allocation supports standard Quality of Service (QoS), cost minimization, energy consumption reduction, increase resource utilization and more over it guaranteed the Service level agreement between the Cloud based IoT system application providers and costumers where user's requirements should be matched in an effective way.

Resource allocation for IoT devices has many challenges due to heterogeneity and distance between the devices. A lot of research work have been done by the industries and academic researchers, but many challenges and issues related to IoT resource allocation are up till now untouched. Many researchers are working to solve these challenges and proposed new methods and algorithms for this. In this paper, the author systematically reviewed many proposed methods for resource allocation in Cloud based IoT environment and classifies these techniques based on characteristics and resource allocation parameters improved by the technique. Furthermore, the author discussed various parameters of IoT resource allocation and what parameters are still needed more improvements.

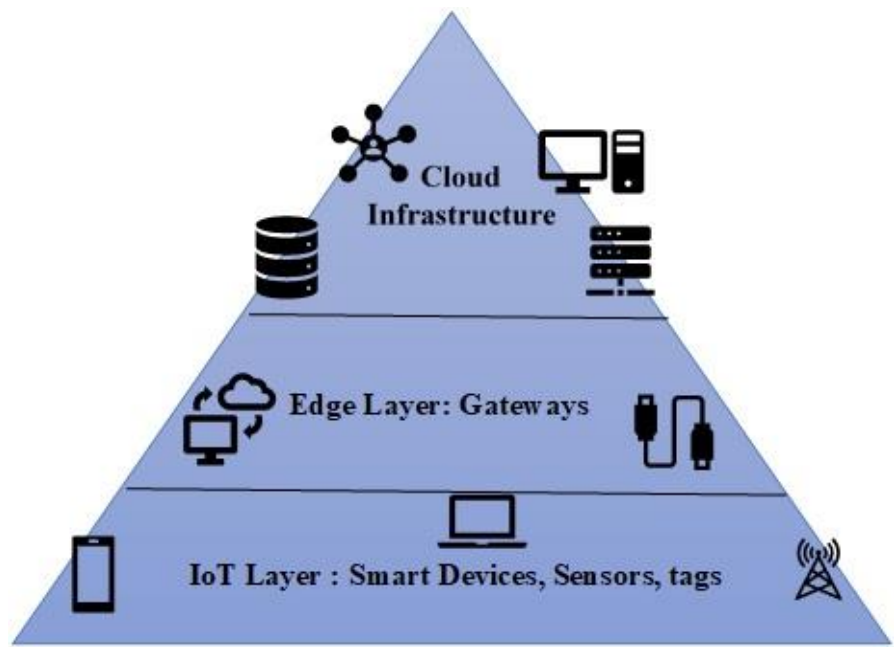

**Fig. 5.** Three tier architecture (IoT, Edge and Cloud)

## 3   Related Works

Many research papers have on been published related to survey and review of IoT technologies, wireless sensor network, integration of IoT and Cloud Computing and IoT architectures. But there is only one survey paper is present on the resource allocation in IoT environment. Many researchers are working in various aspects of IoT architecture. They proposed methods and technologies that support IoT networked system for smooth and cost-effective performance. No review paper exists for Cloud-based IoT resource allocation techniques. The author presented an important survey in [37], that discussed various challenges and issues of IoT resource allocation from the architectural viewpoint. In this review paper author discussed the architectures and infrastructures that are responsible for resource allocation of IoT environment. Resource scheduling and optimization techniques form the Cloud perspectives are not discussed in this paper. Another review paper is [22] based on various IoT technologies and their challenges. In this paper author finds the relation between the fog computing and IoT also discussed various existing resource allocation techniques in fog based IoT environment. The author discussed two main challenges in fog based IoT infrastructure, first is that the fog computing neither care about the nearest node which provides computing resource nor data processing, it only cares about the minimum delay. Second issue the author found is that, the resource allocation between the fog and IoT smart devices, be- cause of limited resource capacity of fog computing. To solve this problem of scarcity of fog resources, the author suggests the use of Cloud Computing above the fog computing. In this paper, the author studied various Cloud-based IoT resource allocation techniques and classified these techniques in different groups. Further limitation and improvement were done in these techniques are also discussed and presented in the tabular form. Various resource allocation parameters have been listed out and defined. The author also discussed How much improvements have been done in these parameters and how much improvement still needed in remaining parameters.

## 4   Classification of Cloud-Based IoT Resource Management Techniques

In this section, the author studied and classified Cloud-based IoT resource allocation techniques in various categories according to the feature provided by the technique un- der study. The detailed classification of IoT resource allocation techniques are described in Figure 6. Various papers have been studied and categorized under these categories of IoT resource allocation.

## 4.1 SLA-Aware IoT Resource Allocation

Service level agreement between the providers and consumer is very important in any service-oriented system. The SLA violation should be minimum to increase the profit of the service provider and satisfied the customer's requirements. Many kinds of research have been done in this area of interest to reduce the SLA violation to improve the system acceptability among the costumers. Some research works are also done related to the SLA oriented resource allocation for IoT enable the system. In [18], the author considered the penalty cost of SLA violation.

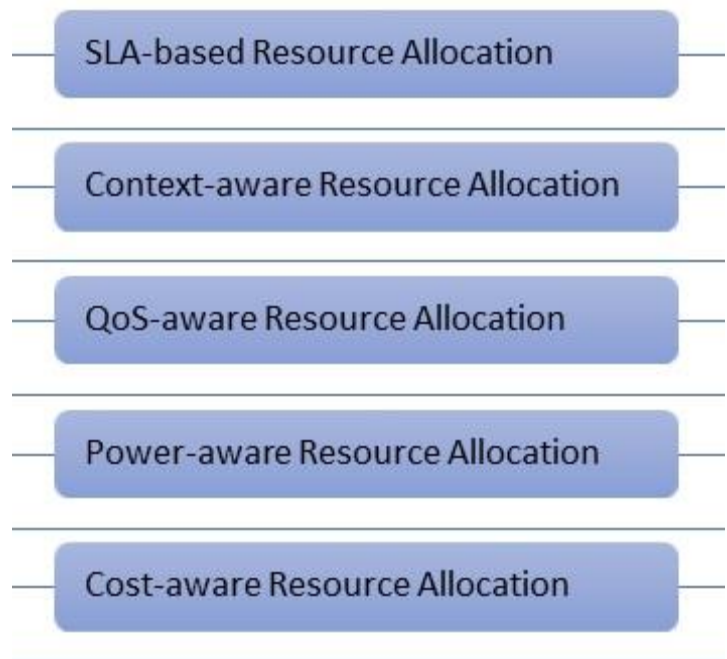

**Fig. 6.** Classification of IoT Resource Allocation Techniques

A combinatorial auction method is proposed to reduce the penalty of SLA violation by calculating the provider's profit and announce a winner to the user who gives maximum profit to the providers so the penalty cost would be minimized to avoid SLA violation. Another SLA violation-based technique is proposed in paper [49]. The author proposed a method to limit the user's task and schedule it efficiently to minimizes SLA violation. The proposed method divides the user task into multiple subtasks and increase the server's capacity to decrease the total measurement time of task execution to reduce SLA violation and maximize providers profits. Another SLA-aware resource allocation method for IoT devices is proposed in [8], the author presented an architectural based service delegation and resource allocation method for Cloud and fog computing based IoT environment. An SLA and QoS oriented algorithm have been proposed that used linearized decision tree for manage user tasks. A new SLA oriented IoT resource allocation problem handled in [50], by buffering scheduling and limit the rate of user's tasks to achieve SLA. The proposed method conforms the SLA violation without knowing the arrival pattern of user's task in advance and works better in huge traffic of IoT devices in the IoT network. These algorithms and their improvement and limitations are described in Table 2.

**Table 2.** SLA-Aware IoT Resource Allocation Techniques

| Algorithm | Improvement | Limitations |
| --- | --- | --- |
| SLA-Based Resource Allocation [18] | Reduce SLA violation and increase system performance | The proposed method is not compared with other techniques |
| Cloud-based SLA-aware Resource Allocation [49] | User's task is divided into sub-tasks to reduce SLA violation | The task arrival time is not calculated dynamically |
| IoT service delegation and resource allocation [8] | Improve system performance and efficiency | There is no evidence of practical implementation. |
| Resource Allocation for IoT applications [50] | Improve System performance and reduce SLA violation by reducing the measurement time of task arrival. | The proposed method is not working in a multi-tenant environment like multiple data centres |

### *4.2 Context-aware IoT Resource Allocation*

Game theory has been applied in many types of research for resource allocation in the device to device communication for IoT devices. Resource allocation is an important aspect for the high performance of data transportation in a wireless network for device to device communication. A location-aware method has been extended in [29], that applied the Nash Equilibrium game model for D2D communication in a cellular network. The author proposed a context-aware algorithm that determines the bandwidth of the network to maximize the total use of each station for communication according to the different situations. A cell association problem has been formulated in [27] as a two-way matching process between the communicating devices. The proposed model used a correlation among the devices that are present in each area network to enhance the cell association technique and guaranteed to a prevalent outcome. Another one-many devices' association based optimal resource allocation method proposed in [2] that increased the resource utilization among the network devices. Dynamic QoS requirements have been achieved by the proposed context-aware resource allocation method between peer to peer IoT network system. Table 3 shows the improvements and limitations of context-aware resource allocation techniques.

**Table 3.** Context-Aware IoT Resource Allocation Techniques

| Algorithm | Improvements | Limitations |
|---|---|---|
| A game-theory based D2D communication in a Cloud-Centric IoT network [29] | D2D communication improved by applied Nash equilibrium game model, maximize bandwidth utilization | This approach is only Cloud-centric and compared with the existing algorithms |
| Correlation-based resource allocation method [27] | Repeated information generated by the different devices has been reduced for better performance in cell association method | The practical evidence is not mentioned in the method |
| Optimal Resource Allocation method [2] | Resource Utilization and performance have been increased | The method has not practical evidence |

## *4.3 QoS-Aware IoT Resource Allocation*

Quality of Service (QoS) is an important aspect of any service-based application. The quality of service should be matched with the service level agreement (SLA). Several researches have been done in this area for a different scenario. A QoS-based IoT resource allocation method has been proposed in [17], which reduced the intrusion between the device to device communication. With the help of PFR method and intrusion limited area control method, the use of resources can be restricted to D2D users. The D2D users get the resources according to the wireless network channel gain, which balances the system workload and enhances the system performance. An analytical model for heterogeneous traffic of M2M devices has been proposed in [48], that used fixed transmission nodes for M2M devices according to the user's requirements to achieve QoS constraints. An optimization protocol has been proposed in [19], which is based on the consensus algorithm for robust and efficient resource allocation in the heterogeneous IoT networks. The algorithm considered the task frequency and buffer occupancy of nodes that are involved in the communication. The method is adaptive in dynamic and heterogeneous feature of IoT device networks. Two communication-based resource allocation method for IoT has been proposed in [20], which used broadcast and gossip methods among the nodes for exchange and update the communication information. The proposed method has been evaluated in three different scenarios: the entire network, single task — single frequency and single task — total frequency. The output error of the system has been reduced to 5% with compared to the centralized solution for the reduction of message transfer and increase the sys-tem reliability. QoS-based resource allocation methods with features and limitations are described in Table 4.

**Table 4.** QoS-Aware IoT Resource Allocation Techniques

| Algorithm | Improvements | Limitations |
| --- | --- | --- |
| Downlink Resource Allocation method [17] | The intrusion among the communication channel has been reduced and enhanced the system performance | Practical implementation is missing |
| Radio Resource Allocation scheme [48] | The traffic among the M2M devices has been reduced | The method is not implemented in real conditions |
| Task Allocation in Group of IoT devices [19] | The optimal resource allocation has been achieved with 5% error | The QoS parameter needs more attention |
| Consensus-based task allocation in IoT [20] | Gossip and broadcast methods are used in which broadcast gives better results | The QoS and real conditions have less considered. |

## 4.4 Energy-Aware IoT Resource Allocation

Due to the huge amount of heterogeneous and more electric power consumed smart devices, it is important to handle the efficient use of electricity or power to reduce the carbon footprints and produce a green computing environment. In many fields like Cloud Computing, fog computing and edge computing, many researchers have been already done a lot of work to reduce power consumption. In IoT based environment there is also some evidence of work in the reduction of power consumption. In [13] a novel resource allocation method for the fog of everything (FoE) has been proposed that presents the energy-delay performance of virtual fog of everything (V-FOE) and compared with the V-D2D technological platform. Three ways of communication between the devices in fog computing architecture network (FOCAN) [45] has been described to meet the QoS and reduce power usage in an effective way. This approach is used in Fog-supported smart cities architecture to share infrastructure resources among smart devices. A generalized Nash equilibrium (GNE) approach and its unique conditions are derived in [3], to handle the heterogeneity of IoT resources, in terms of QoS and resource constraints. A cognitive hierarchy game theory has been used in this method to enable the devices to reach CH equilibrium (CHE) for rationally corresponds to the heterogeneous computing capability and access information of each MTDs and HTDs. The proposed model reduces energy consumption by MTDs by 78%. Another novel ECIoT architecture has been proposed in [36], to additional, improve the system performance by control the process admission and control the power consumption of the resource of the IoT system. A Lyapunov stochastic optimization based cross-layer dynamic network optimization method is used in ECIoT to enhance the system utility. Table 5 shows the limitations and improvements done in these proposed algorithms.

**Table 5.** Energy-Aware IoT Resource Allocation Techniques

| Algorithm | Improvements | Limitations |
|---|---|---|
| Energy-efficient resource allocation for Fog of Everything [13] | Reduce energy consumption and delay and enhance the performance of the FOE system | Real-time testing has not done |
| FOCAN: A Smart city architecture for resource allocation [45] | The method reduces energy consumption and improves latency. | Implementation is not done |
| Cognitive hierarchy theory for resource allocation [3] | The method reduces power consumption by 78% | The only Simulation is done |
| Joint admission control resource allocation [36] | Increased System Throughput and reduce the delay between devices communication | No evidence of practical implementation. |

## *4.5 Cost-Aware IoT Resource Allocation*

IoT network consists of many heterogeneous and many powerful resources needed smart devices which are managed by Cloud Computing, fog computing as well as edge computing infrastructures. These multiple network devices of IoT request the resources for their task completion to fulfill the QoS. The total utilization cost has been calculated according to the served resources and activation cost of each interface devices in the network to serve demand. This cost estimation problem is known Service-to-Interface Assignment cost problem. Two SLA methods are proposed mathematically in [9] to reduce the computational cost. In the first method, the demand for the resource has been fulfilled in one round but in the second method, demand has been fulfilled in multiple rounds. The proposed method splits the activation cost and distributes it in multiple interfaces to reduce the cost of activation. The effective and efficient allocation and release of Cloud resource are important to reduce the service cost of Cloud resource. A multi-agent-based Cloud resource allocation method for IoT devices has been proposed in [39], to audit the dynamic use of resources by the IoT devices. The audit of resource usage help to control the bad resource utilization and increase the utilization of the resources that in turn improve the performance and reduce the total cost of the system. Another method [33], based on Stackelberg game model has been proposed to reduce the network resource cost. The method analysis the lower and upper layer of the net- work and verifies the Nash equilibrium point of the no n-cooperative game between the upper layer of the network to reduce the cost. An iterative method is used to reach the Nash equilibrium by creating the Stackelberg game of the entire network. Multiple heterogeneous network interfaces have been used IoT devices that used a huge amount of services. A resource to heterogeneous service model has been proposed in [10], which is based on mixed-integer linear program (MILP) formulation. The cost of services can be reduced by splitting the services over the different interfaces dynamically. Improvement and limitations of cost-aware resource allocation techniques for IoT environment have been discussed in Table 6.

**Table 6.** Cost-Aware IoT Resource Allocation Techniques

| Algorithm | Improvements | Limitations |
| --- | --- | --- |
| Heterogeneous resource allocation for flexible services [9] | The Cost has been reducing by splitting the services among the interfaces | The cost has been slightly increased in multi-round method |
| Optimizing Cloud resource allocation architecture [39] | Increase system performance, reduce the total cost by minimizing the use of VMs | The cost of ideal resource usage must be reduced |
| Heterogeneous-oriented Resource Allocation Method [33] | Cost resources usage has been reduced | No evidence of practical implementation |
| Flexible Allocation of Heterogeneous Resources [10] | The cost has been reduced and increase the system performance | The more theoretical discussion presents not practically implemented |

## 5 Parameters of IoT Resource Management Techniques

### 5.1 What are the various parameters of resource allocation?

Resource allocation is an important aspect of Cloud-based IoT environment. The devices are connected in the IoT environment by the internet and produced a huge amount of data which are stored in Cloud for further analysis to infer useful information for the organization and application of the system. Many resource allocation parameters have been discussed in [6,7]. In this section, the author defines the various parameters of resource allocation for IoT environment which should be considered for improvement in the development of the IoT resource allocation techniques.

- **Performance** It is an amount of work done by the IoT layer to accomplish the on-demand services of the user's task. Performance of the system should be high.

- **Throughput** The total number of task or job completed by the IoT system is measured as the throughput of the IoT system. Throughput must be high in IoT systems to accomplish all the user's task.

- **QoS** Quality of Service (QoS) is the measurement of the quality services provided by the Cloud system to the end-user which are agreed on the SLA agreement.

- **Delay** Amount of time to respond to a waiting user's task when the system is busy in another task execution. The delay should be minimum in the channel to enhance the system performance.

- **Bit Rate** It an amount of rate at which IoT devices transfer data from one location to another location. Data transfer rate must be high in internet based IoT environment.

- **Reliability** It is the ability to execute the given task in time without affecting by the system failure. Reliability of the system should be high to ensure the user's task completion in time.

- **SLA** Service level agreement between the service provider and the service consumers must be met to overcome the cost overhead and reduce the risk of users and providers relationships violation.

- **Time** It is an important factor in the IoT environment because smart devices produced a huge amount of data regularly. It is a plan to schedule a task in the IoT environment for their execution.

- **Cost** IoT environment uses many services from the Cloud infrastructure providers and in return give money to the service provider. The Cost of services is measured in the performance of the system and how much the system is productive. The cost of the system should be minimum.

- **Energy** IoT environment consists of many network nodes and smart devices which are connected to large data centers. These devices and data centers consumed a huge amount of energy for their proper functioning. The energy consumption should be minimized to reduce cost and make environment carbon footprint-free.

- **Availability** The resource availability is the measure of time and reliability of the re- source in the given period. The availability of the resources should be high to reduce the delay in service.

- **Utilization** Resource utilization is the amount of portion, a resource must be occupied by the system. The resources should be efficiently utilized to enhance the system performance and the reliability of the system.

## 5.2 How much degree of improvement have been done in these parameters?

In this paper, the author studied various Cloud-based IoT resource allocation techniques and classified them into the number of groups. In this section, the parameters of resource allocation which are improved by the given techniques in the classification have been discussed and in Table-7 these parameters are marked corresponding to the method or algorithm under study. From Figure 7, we can observe that how much a parameter has been improved and which one require more attention from the research community.

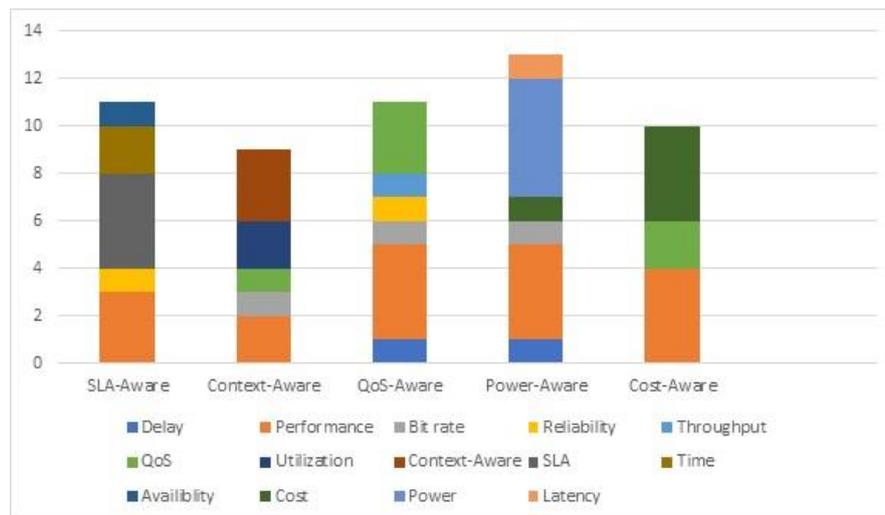

**Fig. 7.** Degree of Improvement in Resource Allocation parameters

## 5.3 How much improvement is required in the remaining parameters?

In the above section, we have already seen that the number of resource allocation parameters have been improved in the proposed algorithms or methods. But still, various resource allocation metrics need improvement and require much attention from the researchers. In Figure 8, various parameters and their corresponding number of algorithms in which they are improved has been shown. From the figure, we can easily identify that the performance of the system has been much improved by many researchers. But the reliability, latency, availability of resources and delay in communication channel have not been improved so much. Therefore, there is a scope for the researchers to propose more resource allocation algorithms that can improve more and more these resource allocation parameters. Multi objective and nature-inspired algorithms can be implemented by the researchers.

## 6  Challenges and Issues

The IoT environment provides potential services to improve the productivity of the Cloud-based IoT ecosystem. Regardless of its potential benefits, there are silently many challenges are present in the Cloud-based IoT resource allocation. In this paper, there are many research papers studied related to IoT resource allocation problem. Most of the papers are based on one or two aspects of the resource allocation, there should be more work needed in the collaboration of IoT with Cloud, Fog and Edge computing for better implementation of IoT ecosystem. Most of the research work done so far has based on simulation and no evidence of implementation of the proposed algorithm for real IoT ecosystem. Real IoT ecosystem has its own challenges and properties and when these algorithms are implemented in real IoT world then there may be some problems occur because of no proper implementation has been tested in real IoT environment. Another challenge in the resource allocation methods studied in this paper is that all aspect of resource allocation like resource discovery, resource modeling, resource provisioning, resource scheduling, resource estimation and resource monitoring has not considered. Optimization in resource allocation techniques is another issue, more of the research only work on the resource allocation technique but optimization in resource allocation is more important for effective and efficient IoT environment.

## 7  Future Directions

IoT environment makes the life of human being so easy. All daily routine work can be done efficiently with the help of IoT devices. Many industries are planning to launch more IoT devices that can make an easier daily task for the human. Many other technologies give life to the IoT infrastructure and provide services. Cloud-based IoT environment for industries application has a future for implementation. The reliability, productivity and cost-effective industry-based applications require more effective and efficient resource allocation. These applications are known as Industries Internet of Things (IIoT) application. Data captured by the smart IoT devices are stored in Cloud and further processed in Cloud infrastructure for inferring knowledge. The amount of data captured by IoT devices is very large that why many researchers proposed Big data as a service to the other industries and Cloud users [30, 31]. Machine learning-based resource allocation has a scope to automate the resource allocation in IoT infrastructure. ML-based resource allocation techniques can overcome the scarcity, over and under- utilization of Cloud resources. Blockchain technology is also an emerging technology which is used for cryptocurrency [23]. Blockchain is a secure method for transferring data from one node to another. It provides privacy and secure communication in a P2P network. Many types of research have been done in IoT and blockchain integration with Cloud Computing [47, 2], but the blockchain is not used in Cloud-based IoT resource allocation.

**Table 7.** IoT Resource Allocation Parameters

| Ref. | Delay | Performance | Bit rate | Reliability | Throughput | QoS | Utilization | Context-Aware | SLA | Time | Availability | Cost | Power | Latency |
|------|-------|-------------|----------|-------------|------------|-----|-------------|---------------|-----|------|--------------|------|-------|---------|
| [18] | - | √ | - | - | - | - | - | - | √ | - | - | - | - | - |
| [49] | - | √ | - | √ | - | - | - | - | √ | √ | √ | - | - | - |
| [8]  | - | √ | - | - | - | - | - | - | √ | - | - | - | - | - |
| [50] | - | - | - | - | - | - | - | - | √ | √ | - | - | - | - |
| [29] | - | √ | - | - | - | - | √ | √ | - | - | - | - | - | - |
| [27] | - | √ | √ | - | - | - | - | √ | - | - | - | - | - | - |
| [2]  | - | - | - | - | - | √ | √ | √ | - | - | - | - | - | - |
| [17] | √ | √ | - | - | - | √ | - | - | - | - | - | - | - | - |
| [48] | - | √ | - | √ | - | √ | - | - | - | - | - | - | - | - |
| [19] | - | √ | √ | - | √ | - | - | - | - | - | - | - | - | - |
| [20] | - | √ | - | - | - | √ | - | - | - | - | - | - | - | - |
| [13] | √ | √ | - | - | - | - | - | - | - | - | - | - | √ | - |
| [45] | - | √ | - | - | - | - | - | - | - | - | - | - | √ | √ |
| [3]  | - | √ | √ | - | - | - | - | - | - | - | - | - | √ | - |
| [36] | - | √ | - | - | - | - | - | - | - | - | - | - | √ | - |
| [9]  | - | √ | - | - | - | √ | - | - | - | - | - | √ | - | - |
| [39] | - | √ | - | - | - | - | - | - | - | - | - | √ | - | - |
| [33] | - | √ | - | - | - | - | - | - | - | - | - | √ | - | - |
| [10] | - | √ | - | - | - | √ | - | - | - | - | - | √ | - | - |

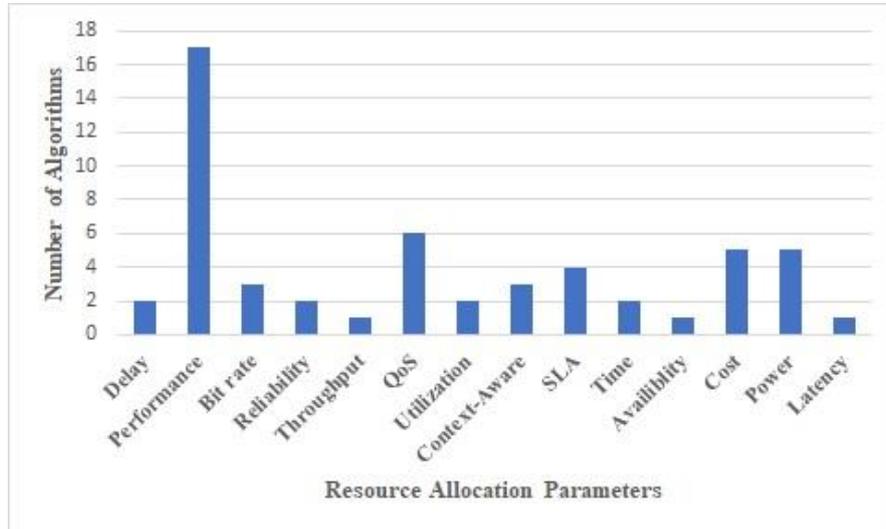

**Fig. 8.** Resource allocation parameters and corresponding number of algorithms

## 8 Conclusion

Resource allocation is an important aspect of Cloud-based IoT environment for effective and efficient working of the novel paradigm. In this paper, Cloud-based resource allocation techniques for IoT environment have been studied and categorized into different groups: SLA-Aware, Context-Aware, QoS-Aware, Power-Aware and Cost-Aware resource allocation. The author systematically reviewed all these techniques and find out the limitations and improvements of these algorithms. Moreover, resource allocation parameters determined for improvements have also listed and defined. A graphical representation of the degree of improvement in these parameters has been shown in the paper. Lastly, challenges and future directions of the study have been discussed.